\documentclass{osa-article}

\journal{oe}

\articletype{Research Article}
\usepackage{siunitx}

\usepackage[acronym, shortcuts]{glossaries}
\newacronym{KOH}{KOH}{potassium hydroxide}
\newacronym{LAE}{LAE}{laser-assisted wet etching}
\newacronym{NA}{NA}{numerical aperture}
\newacronym{CO2}{CO$_2$}{carbon dioxide}
\newacronym{FWHM}{FWHM}{full width at half maximum}
\newacronym{QBB}{QBB}{quasi-Bessel beam}
\newacronym{SD}{$SD$}{standard deviation}

\begin{document}

\title{High-fidelity glass micro-axicons fabricated by laser-assisted wet etching}

\author{Jean-Loup Skora,\authormark{1} Olivier Gaiffe,\authormark{2} Sylwester Bargiel,\authormark{1} Jean-Marc Cote,\authormark{1}, Laurent Tavernier,\authormark{2} Michel de Labachelerie\authormark{1} and Nicolas Passilly\authormark{1,*}}

\address{
\authormark{1}FEMTO-ST Institute, UMR6174 CNRS, Universit\'{e} Bourgogne Franche-Comt\'{e}, F-25000, Besançon, France\\
\authormark{2}Laboratoire de Nanomédecine, Imagerie et Thérapeutiques, EA4662, Centre Hospitalier Universitaire, Universit\'{e} Bourgogne Franche-Comt\'{e}, Besançon, France}
\email{\authormark{*}nicolas.passilly@femto-st.fr} 


\begin{abstract}
We report on the fabrication of micro-axicons made of glass by \ac{LAE} and laser polishing. The employed technique, relying on an efficient direct-writing process by femtosecond laser, allows revealing high fidelity profiles while etched in a heated \ac{KOH} solution. The remaining surface roughness is then smoothened by \ac{CO2} laser polishing. Such polishing is limited to the skin of the component so that the tip is only slightly rounded, with a radius of curvature of nearly \SI{200}{\micro\meter}. It is then shown with 500 µm-diameter axicons that the quasi-Bessel beam is generated closely after the tip, and features a \SI{5.3}{\micro\meter} diameter maintained over a propagation distance of almost \SI{3.5}{\mm}. 
\end{abstract}

\section{Introduction}

\begin{figure}[htp!]
\centering\includegraphics[width=9.15cm]{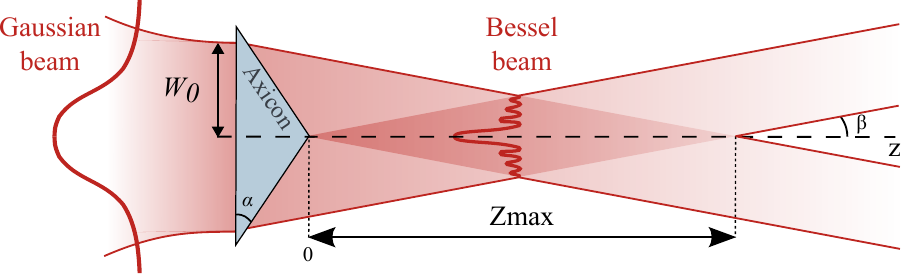}
\caption{Sketch of Bessel beam formation behind an axicon.}
	\label{fig:besselbeamschema}
\end{figure}

Refractive axicons, first proposed in 1954~\cite{mcleod1954}, are conical-shaped continuous profile transmissive components. They are known to produce \ac{QBB}s from an incident collimated beam (Fig.~\ref{fig:besselbeamschema}). Such beams are characterized by an electric-field profile following a Bessel function generated from the interferences superimposition of all deviated beams and diffraction on the conical vertex. This shaping results in a long and narrow focal line along the optical axis ("diffraction-free beam"), rather than a point as in the case of convex lenses. Such non-diffracting unique property makes them useful in applications like alignment and metrology~\cite{hausler1988,davis1996}, second harmonic generation~\cite{piskarskas1999}, 2D micro-fabrication~\cite{li2009}, waveguide writing~\cite{Xin2019}, optical trapping~\cite{garces2002,Suarez2020}, or optical coherence tomography~\cite{Lee2008,lorenser2014} where the associated improved depth of field is an asset. 
To do so, the axicon deviates the incident collimated beam propagating along the z-axis with an angle $\beta$ determined, in paraxial approximation, by the opening angle (also called as base or wedge angle) $\alpha$ of the cone and the refractive index $n$ of the employed material~\cite{mcgloin2005}:

\begin{equation}
\beta = (n-1)\alpha
\label{beta}
\end{equation}

If $ W _{0} $ is the beam-waist of the incident beam, the length of the focal line, $ Z _{max} $, can be approximated to:

\begin{equation}
Z_{max} = W_{0}/\tan\beta
\label{Zmax}
\end{equation}

Whereas the radius $r_{B}$ of the central lobe (first zero) of the \ac{QBB} is described by~\cite{Brzobohaty2008}: 

\begin{equation}
r_{B} = 2.4048/(k\sin\beta) 
\label{rB}
\end{equation}

$r_{B}$ can be then related to the \ac{FWHM} of the central lobe such as \ac{FWHM} $= 1.0532 \times r_{B}$, which is easier to derive experimentally in the ranges where the \ac{QBB} is not properly shaped.

Refractive axicons are usually fabricated by diamond turning~\cite{bass2001handbook} or grinding and polishing~\cite{van2012} of glass or epoxy. In the case of micro-fabrication, most of the employed techniques lead to microstructures made of resist, whose shape can be generated by electron beam lithography~\cite{Cheong2005, Ahluwalia2006}, electric field shaping of SU8 droplet~\cite{cao2013}, interference lithography~\cite{Stankevicius2015}, or 2-photon laser writing~\cite{Schmid2018}. Other means as direct-laser writing of UV-curable epoxy~\cite{huang2014} or chemical etching (positive or negative) to produce axicons at the tip of fibers~\cite{philip2009} have been proposed. Diamond tools are also used to generate molds that are subsequently employed for replication~\cite{Weber2012}. Still, such approaches lead to components in polymer, which are not as robust as if they were made in glass. This can be critical for applications where high laser power has to be handled, e.g. for ablation (microcapillarities, etc.)~\cite{Dudutis2016}, or when the components need to handle harsh conditions. 

Previously, we developed a wafer-level technique based on glass-blowing from multiple silicon-glass cavities to generate micro-axicons~\cite{Carrion2019}. Nevertheless, although the wafer-level character along with the high level of rotational symmetry are undeniable advantages, the technique intrinsically leads to blunt the tip with a radius of curvature reaching few millimeters. 
Recently, axicons made of glass with shorter radii of curvature have been fabricated by femtosecond laser ablation followed by \ac{CO2} laser polishing~\cite{Schwarz2018,Schwarz2020_OPEX,Schwarz2020_JLA,Dudutis2020}. Despite the fact that it is a direct writing method, more adapted to prototyping than large-scale production, it leads to a more conical shape allowing the generation of Bessel beams closer from the tip. However, the significant roughness resulting from the laser ablation requires to perform a strong thermal surface polishing, eventually rounding the tip as well. 

Here, the proposed \ac{LAE} technique is, similarly to laser ablation, a direct writing method, but only for laser-insulating the component's surface so that it is subsequently primarily revealed by wet-etching, especially in potassium hydroxide (\ac{KOH}) aqueous solution, known for anisotropic etching of silicon, rather than glass. However, etch-rate of \ac{KOH} onto the laser insulated part is substantially increased, allowing glass 3-D structuration. Proposed in the early 2000s, such a technique of femtosecond laser pre-irradiation was first employed to enhance etching based on aqueous solution of Hydrofluoric (HF) acid~\cite{Marcinkevicius2001}, but with a much lower selectivity than the one obtained further in \ac{KOH}~\cite{Matsuo2009}. In addition, it has been more focused on microfluidics circuits generation~\cite{Bellouard2004,Sima2018,Liu2021} or microfluidics combined with waveguides for optical sensing~\cite{Osellame2007,Bellini2010}, rather than on micro-optical component fabrication. And when it is employed for this purpose, it is often used to carve a basic geometry which is subsequently shaped by reflow, such as a cube toward a sphere ~\cite{Drs2015} or a disk toward a toroid~\cite{Lin2012} to generate microresonators,  or pillars reflowed in an array of microlenses~\cite{Yang2020}. Even when the primary shape of the lens is drawn from \ac{LAE}, the final seek state, e.g. achieved by flame polishing ~\cite{Qiao2011,Ross2017,Ross2020}, is close to the one minimizing the surface tensions and only slightly modified. 


In the case of an axicon, featuring a much higher spatial frequency in the place of its tip, the laser surface polishing should minimize surface re-shaping. In this framework, we consequently show that the surface roughness resulting from \ac{LAE} is lower than for laser ablation, and then requires lighter thermal surface polishing, eventually better preserving the profile singularities of the component such as the axicon tip. This feature is particularly interesting for micro-axicons since it favors the quasi-Bessel generation in its nearest neighborhood and gives it more potential to be used, e.g., in compact lab-on-chip devices.

\section{Micro-axicon generation}

\subsection{Laser assisted etching}

\begin{figure}[!htp]
	\centering
	\def\svgwidth{.9\columnwidth}
	\graphicspath{}
	\includegraphics[width=13.2cm]{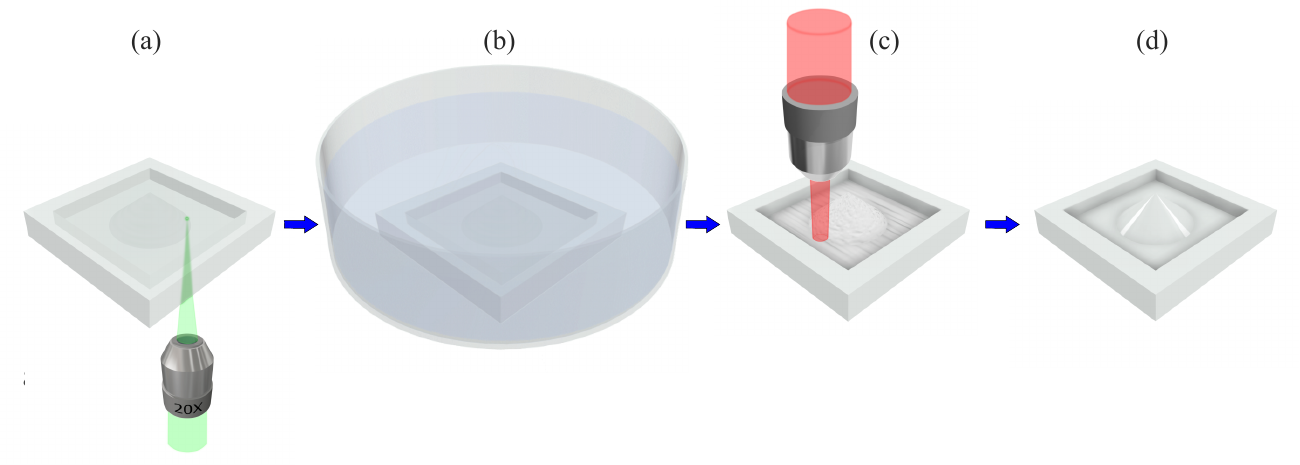}
	\caption{Flowchart of the complete fabrication process of a single micro-axicon in a glass wafer : (a) the axicon outline areas to be subtracted are irradiated by scanning the femtosecond laser-focused spot inside the wafer bulk from its backside,[ along a defined path and starting from the top surface. Here, the irradiated glass is structurally modified and only the refractive index change is visible.] (b) These outline areas are then etched in a hot \ac{KOH} solution leading to the release of the axicon shape. (c) This surface is then smoothed by raster scanning a \SI{150}{\micro\meter} diameter \ac{CO2} laser spot, (d) the resulting axicon surface is optically polished.}
	\label{fig:Flowchart}
\end{figure}

High-fidelity micro-axicons with \SI{500}{\micro\meter} diameter and apex angle of 170$^{\circ}$ ($\alpha$=~5$^{\circ}$) are fabricated in fused silica through a \ac{LAE} manufacturing process~\cite{Marcinkevicius2001} whose direct laser writing step is achieved with a f100 aHead Enhanced system from FEMTOprint (Muzzano, Switzerland). The system consists of a horizontal XY-moving hollow stage holding the glass wafer to be irradiated. The stage stands above a $20\times$ microscope objective, mounted on a vertical Z-translation stage and used to focus inside the glass bulk the beam emitted by a 1030~nm Yb:YAG femtosecond laser (spot size specified at 1.5 µm). The beam pulse energy is set to \SI{230}~nJ and produced at a repetition rate of \SI{1}{\mega\hertz}. The system enables to run programmed sequences synchronizing the laser spot XYZ-displacements with the laser output power in order to pattern glass at a writing speed of \SI{16}~mm/s (Fig.~\ref{fig:Flowchart}(a)). The \ac{LAE} process relies on the structural modification of the glass material resulting from the intense irradiation undergone at the focal point of the femtosecond laser. Since the efficiency of the laser inscription is polarization-sensitive, at least in the femtosecond regime~\cite{LiApplSurfSc2019}, the incident linear polarization is aligned perpendicularly to the writing direction~\cite{HnatovskyOL2005}, in order to optimize the material modification. 

Here, we use \SI{500}{\micro\meter}-thick \SI{20}~$\times$~\SI{20}{\milli\meter} polished fused-silica samples (Corning 7980), although the system can process wafers up to \SI{1}{\milli\meter}-thick and \SI{100}{\milli\meter} diameter. 
The 3D surface of the refractive micro-optical component is raster-drawn with \SI{1}{\micro\meter} line spacing along Y and located few tens of microns below the sample surface. Just before the component's surface, vertical accesses are first insulated. They are aimed to provide evenly distributed access of the etching solution through the sacrificial volume to the 3D laser-exposed surface. By doing so, the sacrificial material does not require to be entirely laser-irradiated but is eventually detached during wet etching~\cite{Butkute2021}. In addition, this scanning strategy prevents from excessive internal stress due to overexposure. The laser irradiation step requires typically \SI{60} minutes for a \SI{500}{\micro\meter}-diameter component. This duration is not proportional to the volume but to the surface of the contour and then is little dependent on the superfluous material, unlike laser ablation for which it has to be entirely laser-treated.

The next step (Fig.~\ref{fig:Flowchart}(b)) consists in a liquid phase etching of the freshly irradiated glass wafer. For this purpose, the glass wafer was immersed into a \SI{10} mol/L \ac{KOH} aqueous solution, maintained at a constant temperature of $85\,^{\circ}$C under sporadic ultrasonic stirring.
It takes approximately \SI{60}~min for the \ac{KOH} solution to react and dissolve the considered areas, i.e. surface and accesses, of the laser-modified glass whatever their number on the wafer. Hence, this etching sub-step allows fast and uniform batch processing. Since these areas were defined as the lens outlines, their dissolution leads to the separation of the undesirable glass parts located above the axicon profile. The laser-modified areas can be etched nearly \SI{900}~times faster than the unmodified glass~\cite{RossOpex2018}, allowing them to be removed with a negligible alteration of the surrounding glass, hence resulting in good machining accuracy. 

Nevertheless, despite dense irradiation paths, the surface of the released micro-axicons shows right after etching a non-negligible roughness that avoids their qualification as optical components, at least for visible wavelengths. Therefore, an additional step concerning the polishing of the surface is required. 

\subsection{$CO_{2}$ laser polishing}

The last fabrication step (Fig.~\ref{fig:Flowchart}(c)) consists in polishing the component's surface through local heating by a \ac{CO2} laser~\cite{Nowak2006,Cao2020}. The Diamond C-30 laser (from Coherent, Santa Clara, CA, USA) emits an infrared beam focused and raster scanned on the surface of the component produced by \ac{LAE}. 

This scanning is done by moving the component with a motorized XY-stage. For each axicon, 13 scans are performed on the X and the Y directions at a speed of \SI{2}{mm/s} with an overlap of \SI{100}{\micro\meter}. The laser power is set at \SI{5.3}~W and the \ac{CO2} laser energy is deposited on a circular Gaussian beam with a \SI{230}{\micro\meter} beam waist (1/e$^2$ spot radius).

Since fused silica is highly absorbing at \SI{10.6}{\micro\meter}, the laser light is absorbed in the vicinity of the lens surface. Therefore, this absorption leads to a significant increase of the glass temperature, limited to a thin layer below the lens surface. As this thin layer temperature increases, its viscosity decreases and eventually reaches a level where the glass is soft enough for surface roughness relaxation effects to occur, which results in a smoothing effect. Indeed, multiple phenomena involving surface tension lead, at first, to the relaxation of high spatial frequencies corresponding to roughness. Since the glass heating is limited to the glass surface, the shape of the lens is rather well-preserved during this process.

\section{Component's profile and surface characterization}

The surfaces of the fabricated axicons were characterized by two different systems: First, an optical profilometer based on white light interferometry (MSA 500 from Polytec, Waldbronn, Germany) was employed to determine surface topography and roughness on a $\SI{886} x~\SI{662} {\micro\meter}^2$ area. Secondly, a mechanical stylus profilometer (Dektak XT from Brüker) was used to measure the profiles of non-polished components, whose roughness is too large for optical profilometer to collect the signal from all the surface points. 

Two batches of components have been fabricated, with a slightly different meshing of the laser-insulated paths. The first tested micro-axicon (MA1), from the first batch, is characterized by a coarser meshing than the second tested one (MA2), taken from the second batch. In addition, those components are compared to a precision axicon (AX125 from Thorlabs) specified at a similar angle ($\alpha = 5\,^{\circ}$) than the fabricated micro-axicons. The various acquired profiles are fitted with the Eq.~\ref{profile}~\cite{depret2002} in order to derive average angles as well as radii of curvature of the axicons tip.

\begin{equation}
e(r) = e_{0}-R_{c}\tan^{2}(\alpha)\sqrt{1+\frac{r^{2}}{R_{c}^{2}\tan^{2}(\alpha)}}
\label{profile}
\end{equation} 

\begin{figure}[h!]
\centering\includegraphics[width=13.2cm]{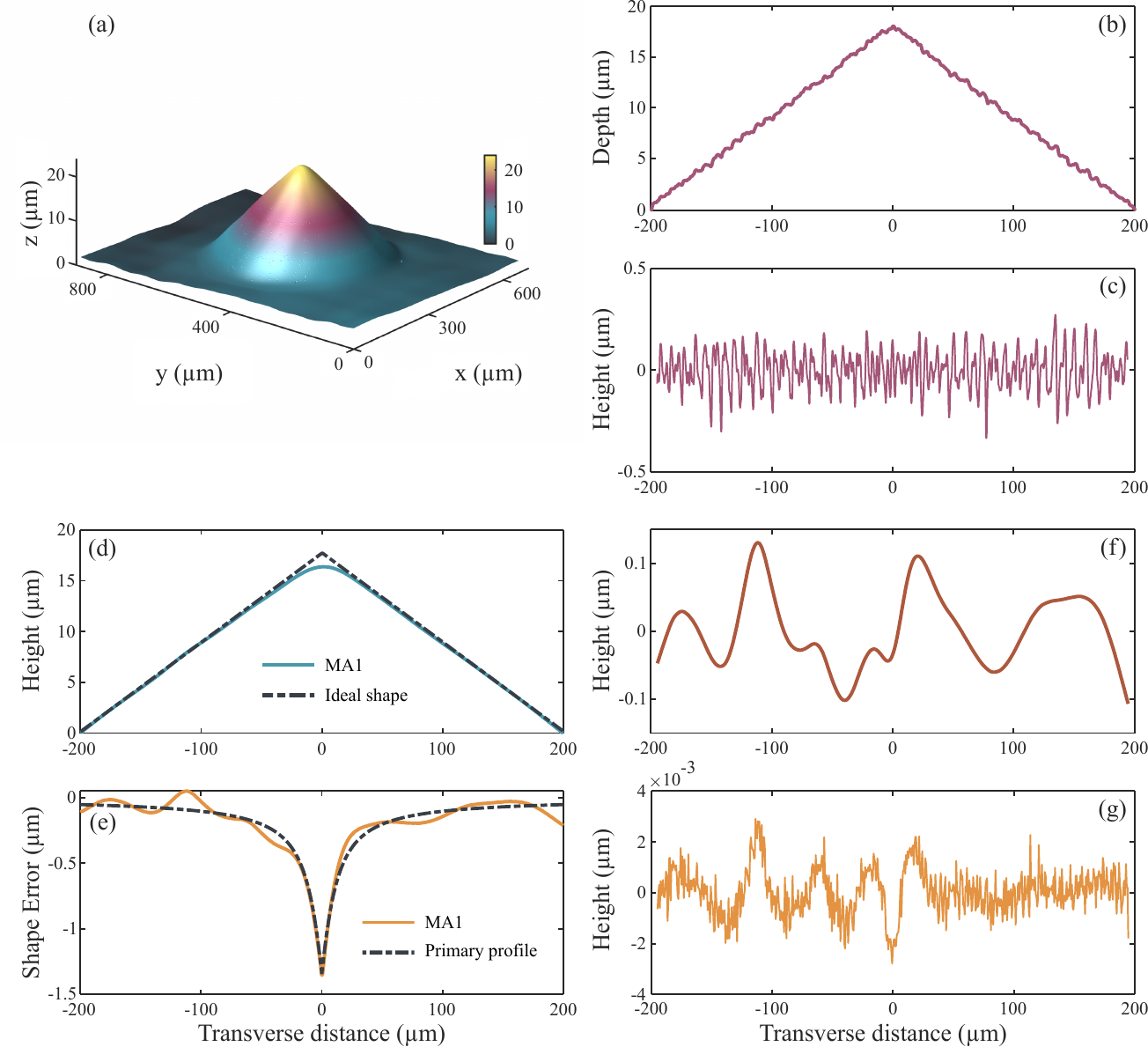}
\caption{Surface profile characterization of the \SI{500}{\micro\meter} - diameter MA1 : (a) 3-D surface relief, (b) Cross-section profile measured right after wet etching. The fitted profile features $R_{c}=$~\SI{28.2}{\micro\meter} and $\alpha = 5.12\,^{\circ}$. (c) Roughness profile from which RMS roughness is derived at the level of $R_{q}=$~\SI{94}{\nano\meter}. (d) Cross-section profile after laser polishing compared to the ideal conical shape (dashed black line). The polished profile now features a more rounded tip with an associated $R_{c}=$~\SI{174}{\micro\meter}. (e) Resulting shape error derived from comparing the two curves of (d). Dashed black line shows the primary profile and allows to derive the RMS waviness in (f) which reaches $W_{q}=$~\SI{54}{\nano\meter}. The roughness of the polished component, extracted by considering a cut-off length of \SI{10}{\micro\meter}, drops significantly. The former rough surface from (c) is  now polished and shown in (g) with a 50x magnification of the ordinate (vertical scale). The calculated roughness is then equal to $R_{q}=$~\SI{0.9}{\nano\meter}. Note that cross sections are mostly derived from white-light profilometry (MSA-500 from Polytec) whereas the rough profiles (b), (c) result from stylus profilometry (Bruker Dektak XT).}
	\label{fig:Axicon_Profile}
\end{figure}


Figure~\ref{fig:Axicon_Profile}(a) reports the topography of one of the \SI{500}{\micro\meter} diameter conical lenses, namely MA1, buried few tens of microns under the wafer surface and measured by white light interferometry.
One of the cross-sections of MA1 is shown in Fig.~\ref{fig:Axicon_Profile}(b) before the polishing step. This rough profile is acquired with the mechanical profilometer equipped with a \SI{2}{\micro\meter} probe. It remains close to the ideal conical shape, with an associated radius of curvature measured around \SI{28}{\micro\meter}. The initial roughness displayed on Fig.~\ref{fig:Axicon_Profile}(c) is at the level of $R_{q}=$~\SI{94}{\nano\meter} evaluated when applying a \SI{10}{\micro\meter} cut-off filter. It can be noted that the roughness observed before polishing varies according to the different axes passing through the axicon tip. Indeed, since the 3D surface of the axicon is generated by writing parallel lines with \SI{1}{\micro\meter} overlap, KOH solution preferentially dissolves and diffuses along the writing lines rather than along the perpendicular direction, resulting in an anisotropic roughness pattern ($R_{q}=$~\SI{45}{\nano\meter} RMS along the writing lines versus $R_{q}=$~\SI{90}{\nano\meter} RMS perpendicularly to them).

The subsequent laser polishing tends to round the tip and thus increase the curvature as shown in Fig.~\ref{fig:Axicon_Profile}(d) where one of MA1's cross-sectional profiles is displayed ($R_{c}=~$\SI{174}{\micro\meter}). The tip rounding is highlighted as well when displaying the shape error (Fig.~\ref{fig:Axicon_Profile}(e), resulting from subtraction between both curves of Fig.~\ref{fig:Axicon_Profile}(d)). Here, it reaches almost \SI{1.4}{\micro\meter} for MA1. Nevertheless, this maximum profile deviation remains almost twice lower than the one measured on the AX125 axicon (\SI{3}{\micro\meter} maximum shape error) and 2 to 4 times lower than the one reported in~\cite{Dudutis2020} measured on laser-ablated components (maximum shape deviation of \SI{3} to \SI{6}{\micro\meter} and associated $R_{c}=$~\SI{600}{\micro\meter}). This higher profile fidelity obtained for the fabricated axicons can be attributed to a lighter surface polishing and an associated higher cut-off of spatial frequencies due to lower initial roughness produced by the \ac{LAE} process. 

From the difference between the shape error profile and the primary profile (dashed black line derived from Eq.~\ref{profile} with a radius set at $R_{c}=$~\SI{207}{\micro\meter}), both displayed on Fig.~\ref{fig:Axicon_Profile}(e), we can subsequently derive the waviness profile (Fig.~\ref{fig:Axicon_Profile}(f)) along with the roughness profile (Fig.~\ref{fig:Axicon_Profile}(g)). The waviness shown here is measured at $W_{q}=$~\SI{54}{\nano\meter}. Hence, \ac{CO2} laser processing leads to a significant decrease of the surface roughness, at the level of $R_{q}=$~\SI{0.9}{\nano\meter} (Fig.~\ref{fig:Axicon_Profile}(g)).

\begin{table}[ht!]
    \small
    \begin{center}
    \caption{
    Representative parameters of axicons' surfaces, derived from 8 radial cross-sections (each$~22.5\,^{\circ}$) collected from surfaces measured with the optical profilometer. Given $\alpha$, $R_{c}$, $W_{q}$ and $R_{q}$ are mean values, averaged from the 8 considered cross-sections.} 
    \label{Tab:Profileparam}
        \setlength{\tabcolsep}{4pt}
    \begin{tabular}{c||cc|cc|cc|cc}
     \hline
         Element & \boldmath$\alpha~(SD^a)$ & Range & \boldmath$R_{c}~(SD)$ & Range & \boldmath$W_{q}~(SD)$ & Range & \boldmath$R_{q}~(SD)$ & Range\\
         & ($^{\circ}$) & ($^{\circ}$) & (\SI{}{\micro\meter}) & (\SI{}{\micro\meter}) & (\SI{}{\nano\meter}) & (\SI{}{\nano\meter}) & (\SI{}{\nano\meter})  & (\SI{}{\nano\meter}) \\
        \hline
        MA1 & 5.08~(0.07) & 5.02~-~5.22 & 191~(37) & 157~-~246 & 68~(13) & 47~-~85 &  0.9~(0.1) & 0.7~-~1  \\
        MA2 & 5.32~(0.10) & 5.20~-~5.45 & 228~(19) & 208~-~258 & 55~(9) & 40~-~71 &  0.8~(0.2) & 0.6~-~1.1   \\
        AX125 & 4.67~(0.01) & 4.66~-~4.67 & 344~(2) & 340~-~346 & 34~(7) & 23~-~43 &  0.7~(0.1) & 0.6~-~0.8  \\
         \hline
         \end{tabular}
         \end{center}
         \footnotesize{$^a$ $SD$ stands for standard deviation}\\
\end{table}

The profile parameters concerning the three considered components are gathered in Tab.~\ref{Tab:Profileparam}. According to the reported values, MA2 appears to be more obtuse than MA1 (by $\approx~0.25\,^{\circ}$) and  differs stronger from the specified $5\,^{\circ}$ than MA1. It is also slightly more asymmetric. Indeed, whereas MA1 wedge angle ranges from  $5.08\,^{\circ}$ (y-axis) to $5.22\,^{\circ}$ (x-axis), it is spread between $5.20\,^{\circ}$ (y-axis) to $5.45\,^{\circ}$ (x-axis) in the case of MA2. This asymmetry is attributed to a higher applied insulation dose, probably consecutive to the finer meshing, resulting also in uneven etch rate. The AX125 mean angle is measured similarly and appears to be more uniform at $4.67\,^{\circ}$ (\ac{SD}$~=~0.01$), value from which a slightly smaller focusing power can however be expected. Note that the angle of the AX125 is evaluated out of topography measurements applied only on the \SI{500}{\micro\meter}-diameter disk centered on the tip, despite its \SI{6.35}{\milli\meter} radius. This can explain the discrepancy with the specified angle of $5\,^{\circ}$.

As reported in Tab.~\ref{Tab:Profileparam}, MA1 shows a higher fidelity around the tip ($R^{\text{~mean}}_{c}=~$\SI{191}{\micro\meter}) than MA2 ($R^{\text{~mean}}_{c}=~$\SI{228}{\micro\meter}) but with a larger \ac{SD} (\SI{37}{\micro\meter} against \SI{19}{\micro\meter}) revealing a more asymmetrical tip. In addition, although MA1 appears to be more asymmetric than MA2 around the tip, it is the contrary concerning their lower part as underlined earlier. Such asymmetries are not experienced with commercial AX125 axicon fabricated by standard grinding and polishing, however, the latter leads to a larger radius of curvature ($R^{\text{~mean}}_{c}=~$\SI{344}{\micro\meter}). 

The surface of the AX125 is also more stretched with nearly twice lower $W^{\text{~mean}}_{q}=$~\SI{34}{\nano\meter} than the one of MA1 ($W^{\text{~mean}}_{q}=$~\SI{68}{\nano\meter}). In addition, the waviness is measured 25\% larger for MA1 than for MA2, as it could be expected from the different meshing of exposed surfaces. Note that, as previously, the wavelength of the applied cut-off filter is set at \SI{10}{\micro\meter}. Finally, the measured surface roughness of the commercial AX125 ($R^{\text{~mean}}_{q}=~$\SI{0.7}{\nano\meter})  is close to its specification given at \SI{0.6}{\nano\meter} and in the same order than the mean surface roughness of MA1 and MA2 measured at \SI{0.9}{\nano\meter} and \SI{0.8}{\nano\meter}, respectively. 

\section{Bessel beam generation}

\begin{figure}[htp!]
\centering\includegraphics[width=13.2cm]{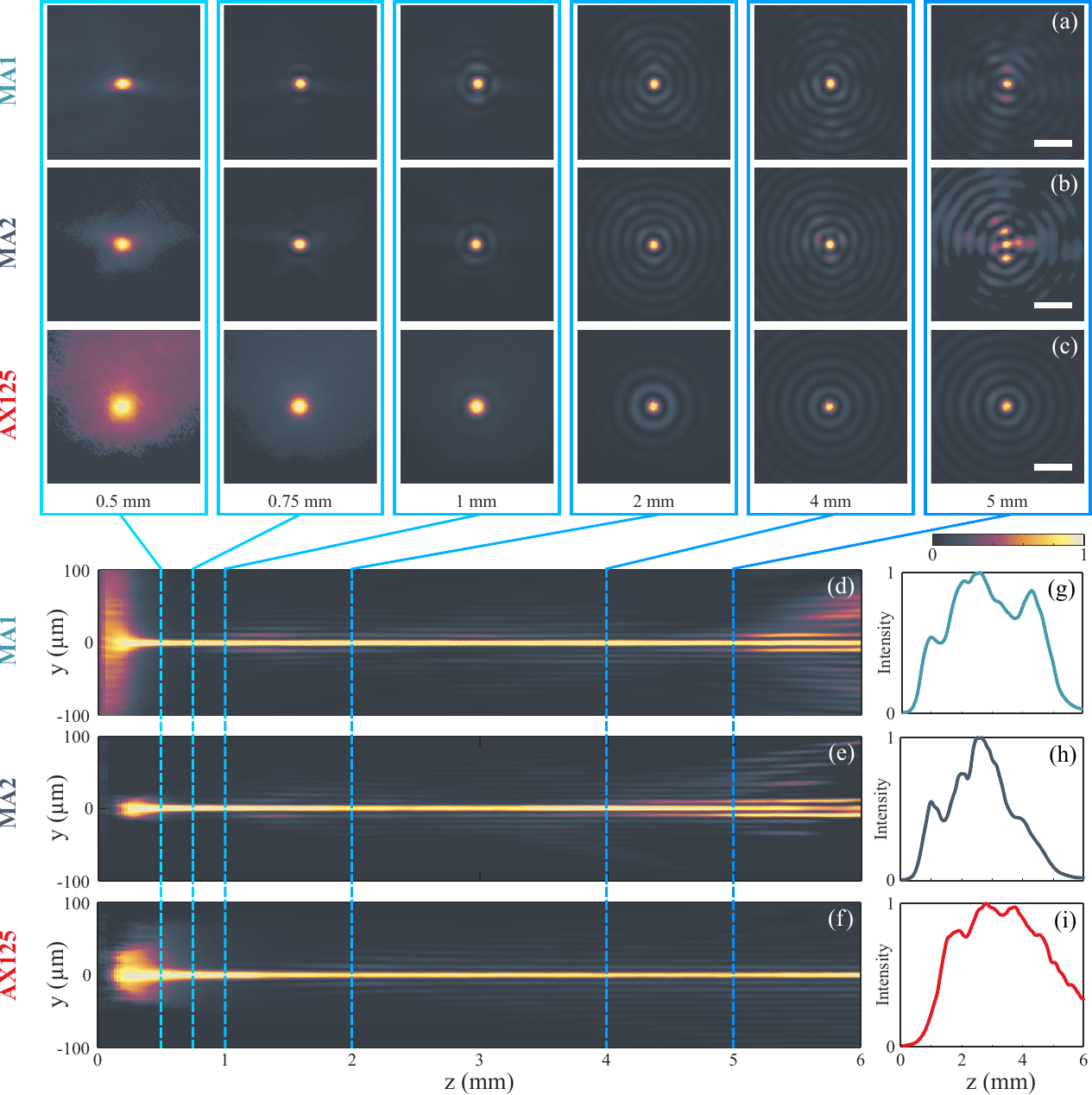}
\caption{Transverse intensity distributions recorded onto the camera at different picked longitudinal distances: Beam shaped by (a) MA1 (b) MA2 (c) commercial AX125 axicon. The white bar is 25 µm long; Associated longitudinally normalized intensity distributions: (d) MA1 (e) MA2 (f) commercial AX125 axicon; On-axis intensity: (g) MA1 (h) MA2 (i) commercial AX125 axicon. }
	\label{fig:transverseplanes}
\end{figure}

Such uncoated components are then employed to generate \ac{QBB}s. Generated intensity distributions are studied with a 3D intensity point spread function (IPSF) measurement system~\cite{baranski2014} where each transverse plane is imaged onto a movable camera so that the whole distribution of the focal volume can be reconstructed. The axicons under test are illuminated with a He-Ne laser ($\lambda=$~\SI{632.8}{\nano\meter}), whose beam is collimated by a beam expander (at $W_0\approx$~\SI{240}{\micro\meter}), whereas the imaging part uses a 10x plan achromat microscope objective from Olympus (NA$~=~0.25$) associated with a tube lens of \SI{150}{\milli\meter} focal length to ensure a magnification of $8.33$. Consequently, each pixel of the CMOS camera (µEye UI-1545LE-M-GL from IDS having $1280~\times~1024$ pixels of \SI{5.2}{\micro\meter} size) corresponds to \SI{0.62}{\micro\meter} in the focal plane. Transverse planes are acquired every \SI{25}{\micro\meter} along the propagation direction thanks to a translation achieved by a servo motor (Ealing 37-1104). 

Figure~\ref{fig:transverseplanes} displays transverse (XY intensity distributions -- Fig.~\ref{fig:transverseplanes}(a) (b) and (c)) and longitudinal (XZ intensity distributions -- Fig.~\ref{fig:transverseplanes}(d) (e) and (f)) sections of the recorded focal volumes. The origin of the longitudinal distance is set at the position of the axicon tip. The Bessel-like intensity distributions, characterized by a sharp and intense central peak surrounded by concentric and evenly separated rings (square of the zeroth-order Bessel function profile) in the transverse planes, and by a non-diffracting feature in the longitudinal plane, are see-able confirming the generation of a \ac{QBB} by the three considered components. The first transverse distributions (the three most to the left) of the beams are noteworthy since they display an earlier shaped QBB for the fabricated micro-axicons MA1 and MA2 than for the commercial AX125, as well as a smaller beam radius, also clearly visible on the longitudinal planes. It can moreover be noticed that such beam shaping is shorter than the one achieved by laser-ablated axicons for which it typically starts at $z~\approx~$\SI{3}{\milli\meter}~\cite{Schwarz2018,Dudutis2020}. The beam central core  early ellipticity of MA1 (Fig.~\ref{fig:transverseplanes}(a): $z~=~$\SI{0.5}{\milli\meter}, $\epsilon~=~1.66$) is attributed to the asymmetry of the component's surface profile around the tip reported in the previous section. On the other side, i.e. at $z~>~$\SI{4}{\milli\meter}, the beam shaped by MA1 (Fig.~\ref{fig:transverseplanes}(a)) does not loose it symmetry as much as the one shaped by MA2 (Fig.~\ref{fig:transverseplanes}(b)), for which asymmetrical sides lobes materialize from $z~>~$\SI{5}{\milli\meter}, in good agreement with the earlier observations of an elliptical contour of the surface profile. 

Nevertheless, as seen on the longitudinal (XZ plane) evolution of MA1 and MA2 between $z~=~$\SI{0.75}{\milli\meter} and $z~=~$\SI{4.75}{\milli\meter} (Fig.~\ref{fig:transverseplanes}(d) and (e)), the central peak remains stable without significant modulation. Note on these graphs that images are normalized (from transverse plane to plane) to better visualize the entire propagation, but the QBB quality can be better appreciated with the on-axis intensity distributions (Fig.\ref{fig:transverseplanes}(g)(h)(i)) where its early generation with respect to the beam shaped by the AX125 is also manifest. 

According to Eq.~\ref{Zmax}, the length of the focal line is influenced by the axicon angle. Taking into account the average wedge angles reported in the Tab.~\ref{Tab:Profileparam}, $Z_{max}$ is equal to \SI{5.92}{\milli\meter}, \SI{5.66}{\milli\meter} and \SI{6.44}{\milli\meter} for MA1, MA2 and the AX125, respectively. The higher angles of the fabricated axicons lead to shortening of their generated focal line length as it can be seen in Fig.~\ref{fig:transverseplanes}(g) and (h) compared to Fig.~\ref{fig:transverseplanes}(i). Additionally, the decline of the on-axis intensity for MA2 (Fig.~\ref{fig:transverseplanes}(h)) is accentuated by the loss of symmetry at the component's contour.

\begin{figure}[htp!]
\centering\includegraphics[width=13.2cm]{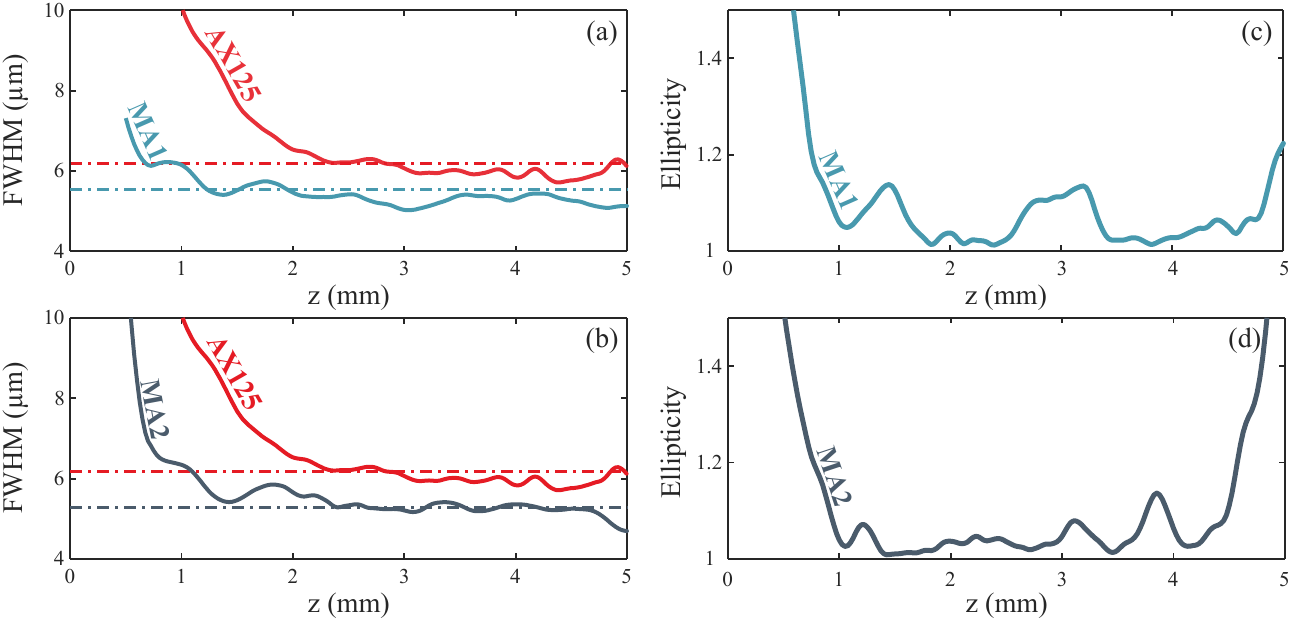}
\caption{(a)-(b) MA1 and MA2 beam central lobe's \ac{FWHM}$_{min}$ derived from experimental IPSF volumes, each being compared to the beam \ac{FWHM}$_{min}$ obtained after the AX125. Dash-dotted lines display the corresponding expected \ac{FWHM}$_{min}$ calculated from approximated Eq.~\ref{rB} of an equivalent ideal (conical) axicon; (c)-(d) Ellipticity of the beam central lobe as a function of the propagation distance for MA1 and MA2.}
	\label{fig:FWHM}
\end{figure}

Finally, the Fig.~\ref{fig:FWHM} displays other features of the shaped beams, in the form of the experimental central lobe's \ac{FWHM}$_{min}$ (Fig.~\ref{fig:FWHM}(a)(b)) and ellipticity (Fig.~\ref{fig:FWHM}(c)(d)) as a function of the longitudinal distance. As expected from Eq.~\ref{rB}, larger angles, such as for MA1 and especially MA2, lead to smaller central lobes. Indeed, whereas the beam generated by the AX125 is characterized by \ac{FWHM}~$\approx~\SI{6}{\micro\meter}$, the ones generated by MA1 and MA2 are closer to~\SI{5}{\micro\meter}. Note that dash-dotted lines display the approximated \ac{FWHM} of an ideal axicon, i.e. fully conical, evaluated according to the Eq.~\ref{rB} and data from Tab.~\ref{Tab:Profileparam} and are plotted at \ac{FWHM}~=~\SI{6.2}{\micro\meter}, \SI{5.5}{\micro\meter} and \SI{5.3}{\micro\meter} for AX125, MA1 and MA2, respectively. As it was already exhibited on the on-axis intensities on Figs.~\ref{fig:transverseplanes}(g)-(h), the \ac{QBB} is generated closer from the tip with MA1 and MA2, i.e. around $z~=~$\SI{1}{\milli\meter}, than with the AX125 for which the \ac{QBB} appears right after $z~=~$\SI{2}{\milli\meter}. This feature, achieved thanks to the reduced radius of curvature of the tip, is particularly interesting for micro-axicons intended to be used within integrated optical microsystems.

From $z~=~$\SI{1}{\milli\meter} to $z~\approx~$\SI{4.5}{\milli\meter}, the ellipticity of the central lobes of MA1 and MA2 remains reasonable ($1.00~<~\epsilon~<~1.15$) (Fig.~\ref{fig:FWHM}(d)-(e)). Taking also into account the on-axis intensities and \ac{FWHM}, this propagation range can be defined as the operational length of the generated \ac{QBB}. It can be noticed that despite a slightly shorter length for the beam shaped by MA2 ($z~\approx~$\SI{4.50}{\milli\meter} instead of $z~\approx~$\SI{4.75}{\milli\meter}), the ellipticity remains close to unity all along its range. This behavior is in good agreement with the previous observations, such as a lower waviness and more symmetrical tip attributed to the finer meshing of the insulated surface of MA2. The latter is then capable of e.g. shaping closely from the axicon tip ($z~=~$\SI{1}{\milli\meter}) until $z~=~$\SI{4.5}{\milli\meter}, a tiny visible \ac{QBB} of \ac{FWHM}~$=~\SI{5.3}{\micro\meter}$, featuring an aspect ratio of 660.

\section{Conclusion}

This paper introduces the use of \ac{LAE}, a rapid prototyping method for the fabrication of micro-optical components made of glass, and in particular micro-axicons employed for \ac{QBB} generation in the visible range. 
It is shown that \ac{LAE}, based on femtosecond laser pre-inscription of glass wafer, followed by wet \ac{KOH} etching, allows building high fidelity refractive surfaces characterized by a reasonable roughness ($R_q~<~$\SI{100}{\nano\meter}). Hence, the required polishing step performed by heating the glass surface with a \ac{CO2} laser faithfully preserves the designed shape of the component whereas decreasing the roughness until less than 1 nm RMS. Thereby, micro-axicons featuring tips with a low radius of curvature ($R_c~<$~\SI{200}{\micro\meter}) are fabricated, leading to the generation of \ac{QBB}s closely from the component's surface (\SI{1}{\milli\meter}). This is particularly interesting for micro-optical components embedded in microsystems that cannot tolerate delayed beam shaping. 

\section*{Funding}
French National Research Agency (ANR) DOCT-VCSEL and µROCS projects (ANR-15-CE19-0012 and ANR-17-CE19-0005), INSERM Robot project (OPE-2017-0123), SMYLE Collegium;

\section*{Acknowledgments}
The authors thank the french RENATECH network and its FEMTO-ST technological facility, Luca Furfaro for fruitfull discussions.

\section*{Disclosures}
The authors declare no conflicts of interest


\bibliography{SKORA_BIB}

\end{document}